\documentclass[twocolumn]{aastex631}

\usepackage{amsmath}
\usepackage{multirow}
\usepackage{makecell}

\usepackage{array}
\newcommand{\head}[2]{\multicolumn{1}{>{\centering\arraybackslash}p{#1}}{{#2}}}

\shorttitle{Radio-selected, optically-detected TDEs discovered in VLASS}
\graphicspath{{./}{figures/}}

\begin{document}

\title{VLASS tidal disruption events with optical flares II: discovery of two TDEs with intermediate width Balmer emission lines and connections to the ambiguous extreme coronal line emitters}

\author[0000-0001-8426-5732]{Jean J. Somalwar}\email{jsomalwa@caltech.edu}
\affil{Cahill Center for Astronomy and Astrophysics, MC\,249-17 California Institute of Technology, Pasadena CA 91125, USA.}

\author[0000-0002-7252-5485]{Vikram Ravi}
\affil{Cahill Center for Astronomy and Astrophysics, MC\,249-17 California Institute of Technology, Pasadena CA 91125, USA.}

\author[0000-0002-1568-7461]{Wenbin Lu}
\affiliation{Department of Astronomy and Theoretical Astrophysics Center, University of California, Berkeley, CA 94720, USA}

\begin{abstract}
The multiwavelength properties of radio-emitting tidal disruption events (TDEs) are poorly understood. In a previous paper, we presented the first sample of radio-selected, optically-detected TDEs, which included two events (VT J1008 and VT J2012) associated with late-time (${\sim}2$ years post-optical flare) intermediate with emission lines that are largely unprecedented from TDEs. In this paper, we investigate these two events in detail. The multiwavelength properties of these events are otherwise consistent with optically-selected TDEs. They are hosted by green valley, E+A/Balmer dominated galaxies with low star formation rates and black holes masses $M_{\rm BH}\approx 10^{5-6}\,M_\odot$. The optical flare shapes are fully consistent with those of optically-selected TDEs, although they are slightly faint and cool at peak. The radio emission from both events is consistent with wide-angle, non-relativistic outflows with $L_R({\rm GHz}) \sim 10^{38}$ erg s$^{-1}$.  Balmer and Helium emission lines are detected from both events with full-width-half-maxima ${\sim}700$ km s$^{-1}$ and asymmetric line profiles. VT J1008 additionally shows coronal line emission with a similar width. The lines from VT J2012 are redshifted by ${\sim}700$ km s$^{-1}$ relative to the host galaxy. We show that these events share many characteristics in common with the ambiguous class of extreme coronal line emitters. We argue that the lines are likely associated with a radiative shock or dense, photoionized clumps of outflowing gas in the circumnuclear medium.
\end{abstract}

\section{Introduction} \label{sec:intro}
Tidal disruption events occur when a star ventures within the tidal radius of a supermassive black hole (SMBH). In the last couple decades, our observational and theoretical understanding of TDEs has dramatically improved, largely thanks to the advent of wide field, time-resolved surveys in the optical and X-ray. Follow-up of optically- and X-ray-selected events at other wavelengths has shed light on the evolution of TDEs but can only provide limited information due to selection biases and limited follow-up resources. 

These limitations have particularly impacted our understanding of the radio emission from TDEs. The physical mechanisms generating TDE radio emission are poorly understood as there have been limited detections of radio emission from TDEs. The landscape of TDE radio emission is beginning to evolve, however, as full sky radio surveys come online.

In a previous paper, we presented the first sample of radio-selected TDEs. This sample included all TDE-like radio transient in VLASS with optical counterparts. In Paper I, we discussed the connections between host galaxy properties/optical flares with the presence of radio emission using a sample of radio-selected TDEs with optical flares. We noticed several possible correlations between multiwavelength properties and radio loudness, including possible trends with optical luminosity and host galaxy SMBH mass. One of the most intriguing discoveries from our sample was the detection of intermediate width Balmer emission from the only two events in our sample in quiescent galaxies. Such emission has not been reported for uniformly selected, bona-fide TDEs before (see \cite{Yao2023arXiv230306523Y} for an event discovered simultaneous with our work): the only previous reports of similar emission are from SDSS-selected flaring galaxies, but these events had poorly sampled multiwavelength data, causing ambiguities about the flare origins.

In Paper I, we presented our full sample. In this paper, we delve into the multiwavelength properties and the interpretation of the two events with intermediate width Balmer emission lines.

\section{Sample Selection}

{\addtolength{\tabcolsep}{-2pt}   \begin{deluxetable}{c | cc}
\centerwidetable
\tablewidth{10pt}
\tablecaption{Properties of our TDEs  \label{tab:host}}
\tablehead{ \colhead{} & \colhead{VT J1008} & \colhead{VT J2012}}
\startdata
R.A. & $10^{\mathrm{h}}08^{\mathrm{m}}53.44^{\mathrm{s}}$ & $20^{\mathrm{h}}12^{\mathrm{m}}29.90^{\mathrm{s}}$ \\
Dec. & $+42^\circ43{}^\prime00.22{}^{\prime\prime}$ & $-17^\circ05{}^\prime56.32{}^{\prime\prime}$ \cr
$\Delta d$ [$\arcsec$] & 0.18 & 0.2 \cr
$z$ & 0.045 & 0.053 \cr
$d_L$ [Mpc] & 205 & 244 \cr
$\log M_{\rm BH}/M_\odot$ & $4.81^{+0.40}_{-0.32}$ & $6.17 \pm 0.31$ \cr
SFR [$M_\odot\,{\rm yr}^{-1}$] & $<1.47$ & $<1$ \cr
$L_\nu($VLASS$)$ [$10^{28}$ erg s$^{-1}$ Hz$^{-1}$] & $9.6\pm0.3$ & $9.9\pm0.9$
\enddata
\centering
\end{deluxetable}}

Our sample selection is described in detail in Paper I; we discuss it briefly here. We identified the two objects described in this work as members of a broader sample of radio-selected TDEs. We compiled this TDE sample from the Very Large Array Sky Survey (VLASS; \citealp{Lacy2020}). VLASS is observing the sky with $\delta > -40^\circ$ at 3 GHz for three epochs with a cadence of ${\sim}2$ years, a per-epoch sensitivity of ${\sim}0.13$ mJy and a spatial resolution of ${\sim}2.5\arcsec$. The first two epochs were completed in 2017/2018 and 2020/2021. The third epoch is ongoing at the time of writing.

We identified TDE candidates using the transient catalog of Dong et al., in prep., who identified all sources that were detected at $>7\sigma$ in E2 but not significantly detected ($<3\sigma$) in E1; i.e., this catalog contains all transients that are rising between E1 and E2. Details about the transient detection algorithm are described in \cite{Somalwar2021} and Dong et al., in prep. We select TDE candidates from this catalog as sources coincident with PanSTARRS-detected galaxies. The galaxies must show no evidence for strong AGN activity in any public archival data. Among the criteria used to identify AGN, we consider: the position of the source on the WISE W1-W2 and W2-W3 color diagram; any evidence for past optical, X-ray, or radio variability/detections that could indicate AGN activity. The host galaxy must also have a photometric redshift $z_{\rm phot}<0.25$, at which redshifts multiwavelength counterparts and host galaxies are more readily detected with reasonable exposure times. The resulting sample has ${\sim}100$ objects.

In Paper I, we focus on the subset of this sample with associated optical flares in data from the Zwicky Transient Facility (ZTF; $gri$ bands) and the Asteroid Terrestrial-impact Last Alert System (ATLAS; $co$ bands). The resulting sample has six objects, three of which are hosted by weak Seyferts, one by a starforming galaxy, and two by quiescent galaxies. The two objects in quiescent galaxies were also detected to have unusual, intermediate width Balmer lines, and thus are the focus of this work. Details of the rest of the sample are presented in Papers I and III. 

The two sources are named using our VLASS transient naming convention: VT J100853.44+424300.22 (VT J1008) and VT J201229.90-170556.32 (VT J2012). We refer to the host galaxies of these objects using the coordinates prefixed with HG (host galaxy; e.g., HG J1008). In plots, we often label the individual transients or host galaxies without the prefixes (e.g., J1008), except when the prefixes are necessary for clarity.

\section{Summary of transient properties}

In this section, we present observations of VT J1008 and J2012 and briefly summarize the multiwavelength properties of each source. We defer discussion of the optical spectral features, which are the main focus of this paper, to the following section.

\subsection{VT J1008}

\begin{figure*}
    \centering
    \includegraphics[width=\textwidth]{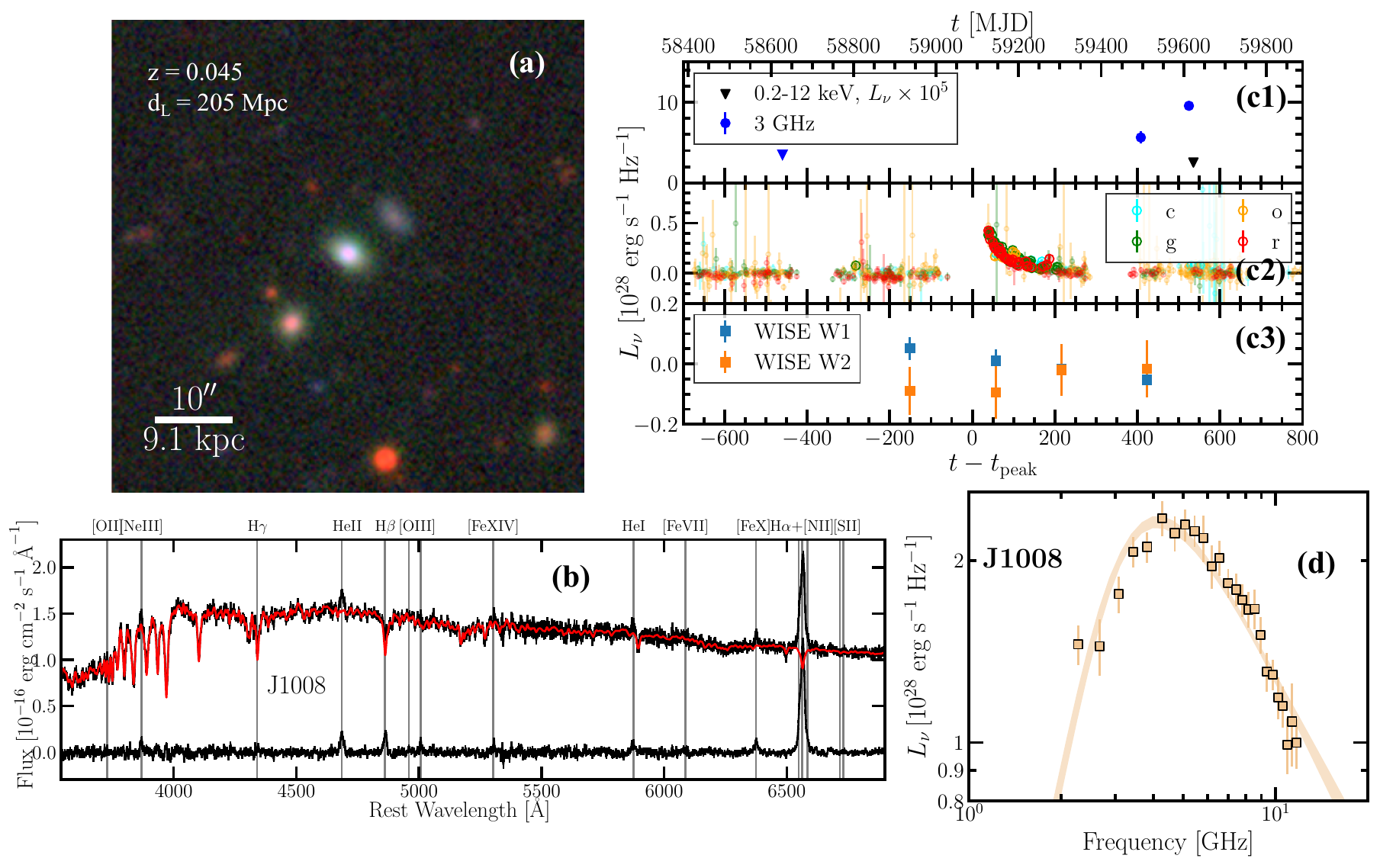}
    \caption{Summary plot for VT J1008. Panel a shows an image of the host galaxy. Panel b shows an example optical spectrum. The observed spectrum is shown on top in black, and the best-fit stellar emission model is shown in red. The observed spectrum with the stellar continuum subtracted is shown on the bottom in black, with the transient emission lines clearly visible. Panel c shows multiwavelength light curves for VT J1008. Panel c1 shows the radio light curve in blue and the X-ray light curve in black. Upper limits are shown as triangles. Panel c2 shows the ATLAS $co$ optical lightcurve. Panel c3 shows the WISE MIR lightcurve, with no obvious flare detected. Panel d shows the radio observations of this source. The radio SED is consistent with a wide-angle, non-relativistic outflow. }
    \label{fig:J1008_sum}
\end{figure*}

We begin our discussion with VT J1008, the multiwavelength properties of which are summarize in Figure~\ref{fig:J1008_sum}. In the following subsections, we break down each panel of Figure~\ref{fig:J1008_sum}.

\subsubsection{Host galaxy properties}

A PanSTARRS optical image of the host galaxy of VT J1008, HG J1008, is shown in the panel a of Figure~\ref{fig:J1008_sum}. It is at a redshift $z=0.045$, or a luminosity distance $d_L = 205$ Mpc. The host galaxy spectrum is shown in panel b. Note that this spectrum is contaminated by transient features. Based on the H$\delta_A$ absorption and an upper limit on the H$\alpha$ EW (we cannot constrain the H$\alpha$ EW exactly because of the transient emission), this galaxy is consistent with being an E+A; i.e., it underwent a starburst in the last ${\sim}$Gyr. There is no strong indication of an AGN, although the faint  [Ne\,III] and [O\,III] emission lines could be produced by a weak AGN. These lines could also be produced by star formation, however. 

We performed an SED fit for HG J1008 in Paper I. We briefly review the results here, but refer the reader to Paper I for a detailed description of our methodology. The host galaxy has a stellar mass $\log M_*/M_\odot = 9.04^{+0.22}_{-0.16}$, and the $3\sigma$ upper limit on the star formation rate is $\log {\rm SFR}/(M_\odot\,{\rm yr}^{-1}) < 1.47$. This host is consistent with the green valley. It is also consistent with being an E+A galaxy.

These SED fit parameters also set constraints on the SMBH masses. From the stellar mass, we can infer a SMBH mass using the \cite{Yao2023arXiv230306523Y} relation for optical TDE hosts: $M_{\rm BH}(M_*) = 4.81^{+0.40}_{-0.32}$. Alternatively, \cite{Yao2023arXiv230306523Y} measured a stellar velocity dispersion for this source $\sigma_* = 44 \pm 3$. Using the $M_{\rm BH}-\sigma_*$ relation from \cite{Kormendy2013}, we find $M_{\rm BH}(\sigma_*) = 5.59 \pm 0.29$, which is consistent with the value measured from the stellar mass. This SMBH mass is small relative to the SMBH masses measured from TDEs by \cite{Yao2023arXiv230306523Y}: VT J1008 has a smaller SMBH mass than 27/32 ($(84 \pm 8)\%$) of their TDEs.

\subsubsection{Optical and IR broadband transient emission}

In panel c2 of Figure~\ref{fig:J1008_sum}, we show the optical lightcurve for VT J1008. This lightcurve was created using the ATLAS\footnote{\url{https://fallingstar-data.com/forcedphot/}} and ZTF\footnote{\url{https://irsa.ipac.caltech.edu/data/ZTF/docs/ztf_forced_photometry.pdf}} forced photometry retrieved using recommended procedures. Both surveys detected an optical flare from this source starting on MJD${\sim}59100$. The flare rise was missed, but the decay was well-sampled. \cite{Yao2023arXiv230306523Y} fit this lightcurve as an evolving blackbody with the temperature fixed to that at peak luminosity. They found a peak blackbody luminosity of $\log L_{\rm bb}/({\rm erg\,s}^{-1}) = 42.98$ and peak temperature $\log T_{\rm bb}/{\rm K} = 4.15$. The radius at peak luminosity was $\log R_{\rm bb}/{\rm cm} = 14.76$. The time to rise from half-max-luminosity to max-luminosity was $t_{\rm rise, 1/2} = 11.8^{+1.5}_{-1.3}$ days, and the time to decay from max to half-max was $t_{\rm decay, 1/2} = 23.1^{+1.8}_{-1.2}$ days. This peak temperature is cool for a typical optically-selected TDE: it is cooler than 25/32 ($(78\pm9)\%$) of the optical TDEs in \cite{Yao2023arXiv230306523Y}. The peak luminosity is lower than every TDE in the \cite{Yao2023arXiv230306523Y} sample. The rise and decay times are typical of optically-selected TDEs.

During the writing of this work, VT J1008 rebrightened in the optical and has been identified as a repeating partial TDE. Although it is plausible that the repeating nature of this source has affected the emission, we do not consider the rebrightening here and refer the reader to \cite{ptde} for details.

In panel c3, we show the IR lightcurve for this source from the NEOWISE survey, processed using the methods described in \cite{Somalwar2023ApJ...945..142S}. We do not see any significant IR variability.

\subsubsection{X-ray emission}

We checked public X-ray survey data, including the XMM Newton Slew Survey, etc, for archival detections of VT J1008. No detections were reported. The pre-optical-flare X-ray upper limit was $10^{42.9}$ erg s$^{-1}$. The tightest post-optical flare limit is from our Swift/XRT ToO. We obtained X-ray observations of VT J1008 on MJD 59638 using a 3.5 ks exposure with the Swift/XRT telescope (PI Somalwar). The source was not detected, with a $3\sigma$ luminosity upper limit of $10^{41.8}$ erg s$^{-1}$. This upper limit is shown in panel c1 of Figure~\ref{fig:J1008_sum}.

\subsubsection{Radio emission}

The radio lightcurve for VT J1008 is shown in panel c1. VT J1008 was first observed by VLASS on MJD 58628; this nondetection was ${\sim}460$ days before the optical peak. It was observed again on MJD 59496 ($524$ days post-optical peak) and was detected as a $1.13 \pm 0.15$ mJy source, corresponding to a 3 GHz luminosity of $L_\nu = (5.6 \pm 0.7)\times 10^{28}$ erg s$^{-1}$ Hz$^{-1}$. We observed the source with the VLA on MJD 59612 ($525$ days post-peak) in the CLSX bands. The 3 GHz radio luminosity from this SED had risen since to VLASS observation to $(9.6 \pm 0.3) \times 10^{28}$ erg s$^{-1}$ Hz$^{-1}$. If we assume that the radio-emitting outflow was launched at optical peak, this rise corresponds to a $L_\nu \propto t^{2.1^{+0.6}_{-0.5}}$ power-law evolution, which is consistent with expectations for a constant velocity outflow.

The VLA radio SED is shown in panel d. We modelled the SED as a synchrotron outflow following the methodology described in Appendix E of Paper I, where we assume a spherically symmetric outflow with magnetic field $B$, electron density $N_0$, and radius $R$. We assume the electrons have a power-law energy distribution with index $\gamma$. We assume equipartition with $\epsilon_E = 11/17$ and $\epsilon_B = 6/17$, corresponding to the minimum energy solution. The total energy in the outflow is given by $E$. We find a radius $\log R/{\rm cm} = 17.06^{+0.01}_{-0.01}$, a magnetic field $\log B/{\rm G} = -0.73^{+0.05}_{-0.04}$, an electron density $\log N_0/{\rm cm^{-3}} = 3.5^{+0.2}_{-0.1}$, and an energy $\log E/{\rm erg} = 49.4^{+0.1}_{-0.1}$. Assuming the outflow was launched at the optical peak with velocity $v$, the best-fit radius gives $\log \beta = \log v/c = -1.37^{+0.1}_{-0.1}$, or $v \approx 1.3\times10^3$ km s$^{-1}$. 

The radio SED is thus consistent with a non-relativistic, wide-angle outflow launched near optical peak. 

\subsection{VT J2012}

\begin{figure*}
    \centering
    \includegraphics[width=\textwidth]{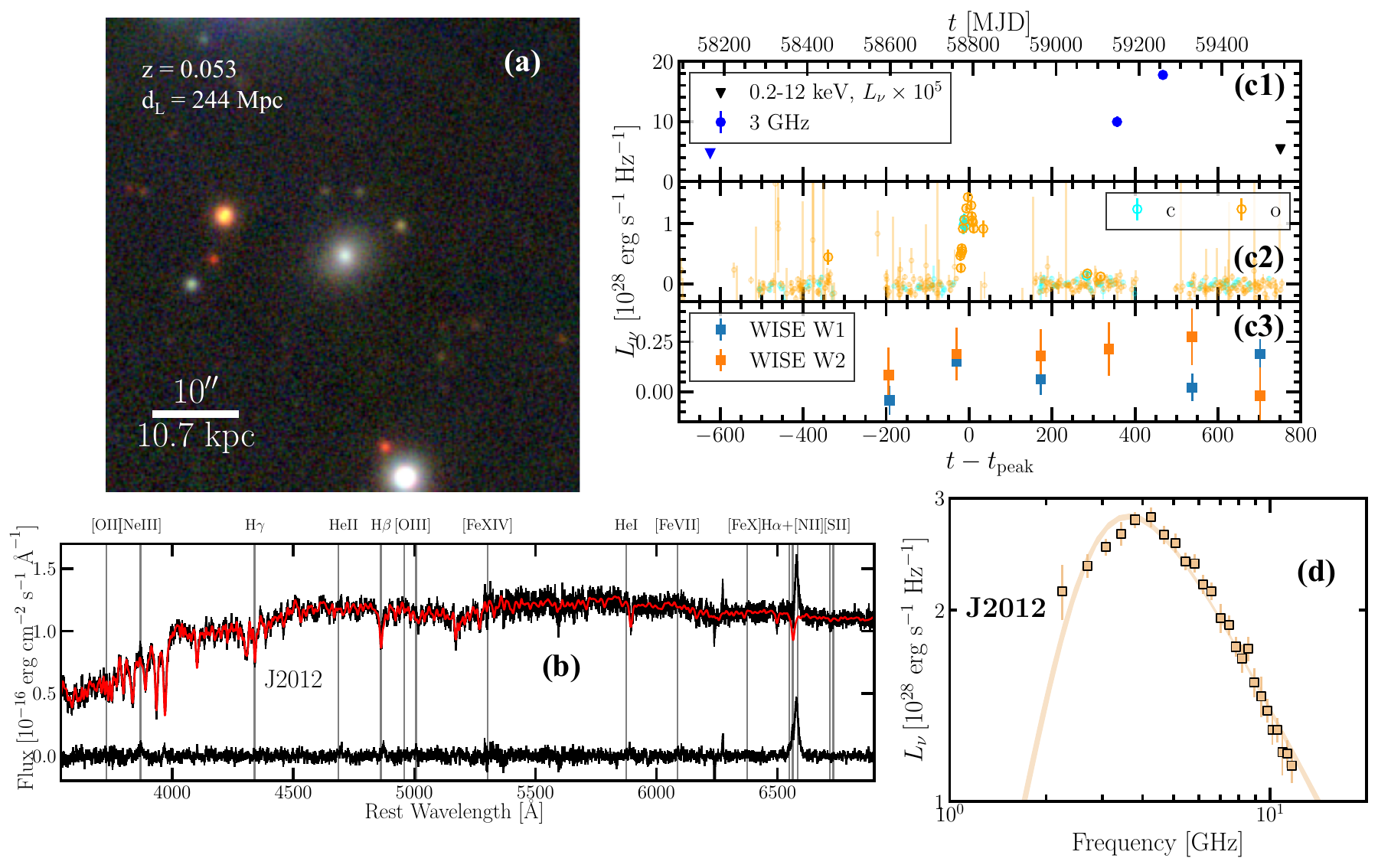}
    \caption{Summary plot for VT J2012, in the same format as Figure~\ref{fig:J1008_sum}.}
    \label{fig:J2012_sum}
\end{figure*}

\subsubsection{Host galaxy properties}

An optical image from the DECam Legacy Survey of HG J2012 is shown in the panel (a) of Figure~\ref{fig:J2012_sum}. It is at a redshift $z=0.053$, or a luminosity distance $d_L = 244$ Mpc. The host galaxy spectrum (contaminated by transient features) is shown in panel (b). From the H$\delta_A$ absorption and an upper limit on the H$\alpha$ EW, this galaxy is a balmer strong galaxy, which have slightly older stellar population than E+A galaxies. There is no indication of an AGN, although faint [Ne\,III] emission may be present. 

From an SED fit performed in Paper I, the host galaxy has a stellar mass $\log M_*/M_\odot = 9.90_{0.09}^{0.32}$. The $3\sigma$ upper limit on the star-formation rate is $\log {\rm SFR}/(M_\odot\,{\rm yr}^{-1}) < 1$.  This host is consistent with the green valley.

From the stellar mass, we can infer a SMBH mass using the \cite{Yao2023arXiv230306523Y} relation for optical TDE hosts: $M_{\rm BH}(M_*) = 6.55^{+0.24}_{-0.32}$. In paper I, we measured a stellar velocity dispersion for this source $\sigma_* = 59 \pm 2$. Using the $M_{\rm BH}-\sigma_*$ relation from \cite{Kormendy2013}, we find $M_{\rm BH}(\sigma_*) = 6.17 \pm 0.31$, which is consistent with the value measured from the stellar mass. This SMBH mass is near the median of the SMBH masses from \cite{Yao2023arXiv230306523Y}

\subsubsection{Optical and IR broadband transient emission}

The optical lightcurve for VT J2012 is shown in panel c2 of Figure~\ref{fig:J2012_sum}. This source was only detected by the ATLAS survey, and the lightcurve was created using recommended procedures. The flare was first detected on MJD${\sim}59100$. Paper I fit this lightcurve to the same model as used for VT J1008, and found a peak blackbody luminosity of $\log L_{\rm bb}/({\rm erg\,s}^{-1}) = 43.07$ and peak temperature $\log T_{\rm bb}/{\rm K} = 3.93$. The radius at peak luminosity was $\log R_{\rm bb}/{\rm cm} = 15.26$. The time to rise from half-max-luminosity to max-luminosity was $t_{\rm rise, 1/2} = 10.2^{+1.5}_{-1.1}$ days, and the time to decay from max to half-max was $t_{\rm decay, 1/2} = 15^{+14}_{-10}$ days. Like VT J1008, this source shows a remarkably cool, faint flare relative to the \cite{Yao2023arXiv230306523Y} sample.

In panel c3, we show the IR lightcurve for this source from the NEOWISE survey, processed using the methods described in \cite{Somalwar2023ApJ...945..142S}. Like for VT J1008, We do not see any significant IR variability. 

\subsubsection{X-ray emission}

We checked public X-ray survey data, including the XMM Newton Slew Survey, etc, for archival detections of VT J1008. No detections were reported. The pre-optical-flare X-ray upper limit was $10^{43.1}$ erg s$^{-1}$. The tightest post-optical flare limit is from our Swift/XRT ToO. We obtained X-ray observations of VT J2012 on MJD 59536 using a 1.6 ks exposure with the Swift/XRT telescope (PI Somalwar). The source was not detected, with a $3\sigma$ luminosity upper limit of $10^{42.1}$ erg s$^{-1}$. This upper limit is shown in panel c1 of Figure~\ref{fig:J1008_sum}.

\subsubsection{Radio emission}

The radio lightcurve for VT J2012 is shown in panel c1. VT J2012 was first observed by VLASS on MJD 58166 (${\sim}624$ days pre-peak) and the luminosity upper limit was $L_\nu = 4.7\times10^{28}$ erg s$^{-1}$ Hz$^{-1}$. It was observed again on MJD 59147 ($357$ days post-optical peak) and was detected as a $(9.9 \pm 0.9) \times 10^{28}$ erg s$^{-1}$ Hz$^{-1}$ source ($1.48\pm0.13$). We observed the source with the VLA on MJD 59258 ($467$ days post-peak) in the CLSX bands, and the 3 GHz radio luminosity had risen to $(14.9 \pm 0.2) \times 10^{28}$ erg s$^{-1}$ Hz$^{-1}$. This corresponds to a $L_\nu \propto t^{1.5^{+0.4}_{-0.3}}$ power-law, which is consistent with a constant velocity.

The VLA radio SED is shown in panel d. We modelled the SED as a synchrotron outflow following the same methodology used for VT J1008. We find a radius $\log R/{\rm cm} = 17.225_{-0.006}^{+0.006}$, a magnetic field $\log B/{\rm G} = -0.83_{-0.02}^{+0.02}$, an electron density $\log N_0/{\rm cm^{-3}} = 3.27_{-0.08}^{+0.08}$, and an energy $\log E/{\rm erg} = 49.69_{-0.05}^{+0.05}$. Assuming the outflow was launched at the optical peak with velocity $v$, the best-fit radius gives $\log \beta = \log v/c = -1.178_{-0.006}^{+0.006}$, or $v \approx 2 \times 10^4$ km s$^{-1}$.

Like that of VT J1008, this radio SED is consistent with a non-relativistic, wide-angle outflow launched near optical peak. 

\subsection{Summary}

We conclude with a brief summary of the results from this section:
\begin{itemize}
    \item VT J1008 and VT J2012 are hosted by quiescent galaxies with no evidence for strong AGN activity. The galaxies are both E+A or Balmer-strong galaxies in the green valley. Both galaxies have SMBH masses $M_{\rm BH} \approx 10^{5-6}\,M_\odot$. 
    \item VT J1008 and VT J2012 have optical counterparts with peak blackbody luminosities $L_{\rm bb} \approx 10^{43}$ erg s$^{-1}$ and temperature at peak $T_{\rm bb} \approx 10^4\,K$, which are cooler and fainter than those of typical optically-selected TDEs. The rise and decay times of the optical flares are typical of optically-selected TDEs.
    \item Neither VT J1008 or VT J2012 have IR or X-ray counterparts. 
    \item VT J1008 and VT J1008 both are associated with radio transients that turned on ${\sim}1-2$ years post-optical peak. They both have 3 GHz luminosities ${\sim}10^{29}$ erg s$^{-1}$ Hz$^{-1}$, which is a typical luminosity for optically-selected, radio-detected TDEs. The radio SEDs of both sources are consistent with low velocity ${\sim}10^{-1}c$, wide angle outflows with energies ${\sim}10^{49.5}$ erg.
\end{itemize}

\section{The transient optical spectral features} \label{sec:lines}

While keeping the broader picture of the multiwavelength properties of our two radio TDEs in mind, we now delve into the intermediate width transient lines, which are the focus of this work. Zoom-ins on these features are shown in Figure~\ref{fig:line_fits}. In this section, we present our methodology for constraining the emission line properties and present the results. We then briefly compare the observed lines to other transient observations, but defer a detailed discussion of the origins of the lines to Section~\ref{sec:discussion}.

\subsection{Methodology}

\begin{figure*}
    \centering
    \includegraphics[width=\textwidth]{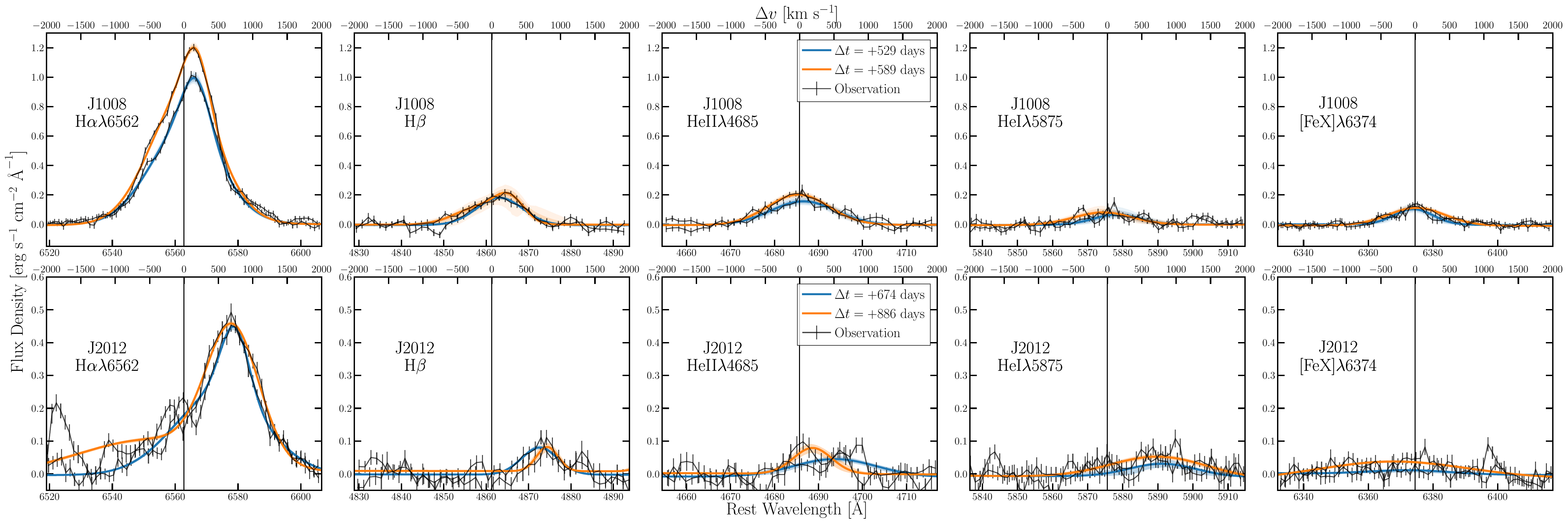}
    \caption{Zoom-ins on select optical lines from the low resolution LRIS observations of VT J1008 and VT J2012. The top row shows observations of VT J1008 and the bottom row shows observations of VT J2012. The observations are shown in black, and the best-fit models are shown as colored lines. The colored bands denote $1\sigma$ uncertainties. The blue fits correspond to the first observation epochs, and the orange fits correspond to the second epochs. The features blueward of H$\alpha$ in the J2012 panel are caused by telluric features}
    \label{fig:line_fits}
\end{figure*}

We first constrain the properties of the transient spectral lines. We use two sets of observations, which we process separately. 

First, we considered low resolution but flux calibrated observations with the Low Resolution Imaging Spectrometer (LRIS) on the Keck I telescope. We observed VT J1008 on MJD 59676 for 20 min{.} using the $1\arcsec$ slit with the standard star Feige 34 and on MJD 59616 for 10 min{.} using the standard star G191-B2B. We centered all observations on the galactic nuclei using a parallactic angle. We used the 400/3400 grism, the 400/8500 grating with central wavelength 7830, and the 560 dichroic. The resulting wavelength range was ${\sim}1300{-}10000\,{\rm \AA}$ and the resolution $R{\sim}700$. We observed VT J2012 on MJD 59464 for 10 min using the $1\farcs0$ slit with the standard stars BD+28 and G191-B2B for the blue/red sides, respectively, and on MJD 59678 using the standard star Feige 34. We reduced all spectra using the \texttt{lpipe} code with standard settings.

To constrain the properties of the transient spectral lines, we first must remove the host galaxy stellar emission from all the spectra. We fit each spectrum with the \texttt{ppxf} full spectrum fitting tool using the MILES templates \citep[][]{Vazdekis2010} following the method detailed in Appendix B of \cite{Somalwar2021}. The resulting best-fit stellar component is shown, for each source, in red in the panel b's of Figure~\ref{fig:J1008_sum} and Figure~\ref{fig:J2012_sum}. We then create a nebular spectrum by subtracting the stellar component from the galaxy spectrum, and the result is shown at the bottom of the panel b's of Figure~\ref{fig:J1008_sum} and Figure~\ref{fig:J2012_sum}.

With these low resolution emission line spectra in hand, we now can fit the line profiles. We consider the following lines, for reasons that will become clear later in this work: H$\alpha$, H$\beta$, [O\,III]$\lambda 5007$, [O\,III]$\lambda 4959$,  [O\,II]$\lambda\lambda 3727,3729$, and He\,II$\lambda 4686$. We first fit each line to a Gaussian with free centroid, width, and flux. We require the centroid be within $2000$ km s$^{-1}$ of the expected wavelength given the host redshift. We set the lower bound of the width to be such that the line FWHM is greater than $6{\rm \AA}$, corresponding to a rough lower limit on the LRIS resolution. The width upper bound is $2000$ km s$^{-1}$, which does not affect the fits. We adopt broad, uninformative priors for the flux: $f \in [0, 10^{-13}]$ erg cm$^{-2}$ s$^{-1}$. For the [O\,II]$\lambda\lambda 3727,3729$ complex, we fit a single Gaussian rather than two because we do not expect the doublet to be resolvable given the large LRIS FWHM.

For some of the Balmer lines, this single Gaussian fit produces a statistically inconsistent $\chi^2$. In those cases, we run a third fit with two Gaussian components, each with independent widths, amplitudes, and centroids. For J2012, we fix the centroid of one component at the host redshift. This produces a consistent $\chi^2$ in each case.

From both events, the [O\,II] doublet is not significantly detected, and the [O\,III] line is detected with a narrow width ${\sim}100$ km s$^{-1}$. Given the narrow widths of the [O\,III] emission, we do not believe it is coming from the same location as the Balmer and Helium emission. Instead, it is consistent with being stable host galaxy emission. To constrain the presence of intermediate width, transient emission at the location of these oxygen lines, we repeat the Gaussian fit but fix the width to FWHM$\approx700$ km s$^{-1}$. In the case of ${[\rm O\,III}]\,\lambda 5007$, we do not subtract out the narrow line component before this fit, so the broad fit absorbs the flux from this narrow line. We choose to do this because it produces the most conservative upper limit on the broad line flux for this emission line, even though the resulting fit has a high $\chi^2$. Subtracting out the narrow component would not affect our conclusions; it would simply tighten the bound on the presence of transient ${[\rm O\,III}]\,\lambda 5007$ emission. 

The resulting best-fit parameters of the lines from the low resolution spectra are listed in Table~\ref{tab:lines}, and the fits are shown in Figure~\ref{fig:line_fits}.

In addition to these low resolution spectra, we use high resolution spectra that are not flux calibrated. The high resolution spectra are taken at later times than the low resolution spectra, so they provide constraints on the line profiles over a longer timescale. Moreover, while we cannot set luminosity constraints using the high resolution spectra, we can study the line profiles in detail; in particular, we can distinguish between multiple, blended lines or a single line with a broad profile. We obtained spectra of these obejcts with the Echellette Spectrograph and Imager (ESI) on the Keck II telescope. We used the Echelle mode for all observations. We observed VT J1008 on MJD 59908 for 22.5 min using the $0\farcs3$ slit and the standard BD+28, and on MJD 60029 for 45 min using the $0\farcs5$ slit and the standard BD+33. We observed VT J2012 on MJD 59876 for 20 min using the $0\farcs5$ arcsec slit and the standard BD+28. We reduced the observations using the \texttt{makee} code with standard settings, and removed telluric lines using recommended procedures. For both objects, the H$\alpha$ lines fall in a region with strong telluric absorption and bright sky lines. Fortunately, the spectra are of sufficiently high resolution that we can still study the smooth line profiles, but all spikes and other unusual features in the data are due to this contamination.

\begin{figure*}
    \centering
    \includegraphics[width=\textwidth]{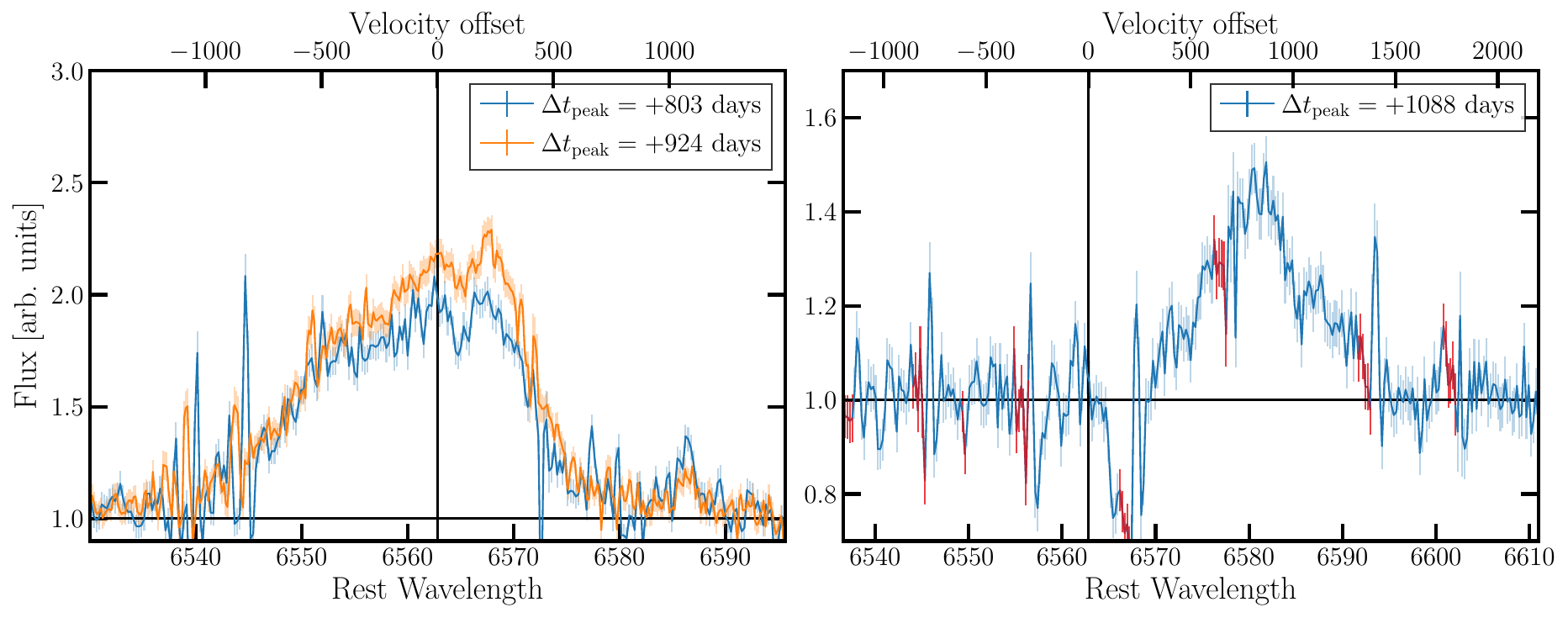}
    \caption{H$\alpha$ spectral profiles from the medium resolution ESI observations. The observations of VT J1008 are in the {\it left} panel, and those of VT J2012 are in the {\it right} panel. The blue line shows the first epoch of observations and the orange line shows the second epoch. The flux for each epoch is normalized to the local continuum, which is not expected to be the same in both observations. In the {\it right} panel, regions particularly impacted by strong sky lines are shown in red. }
    \label{fig:ESI_Halpha}
\end{figure*}

Zoom-ins on the resulting H$\alpha$ line profiles are shown in Figure~\ref{fig:ESI_Halpha}. While other lines are detected (VT J1008: [O\,III]$\lambda 5007$, He\,II$\lambda 4686$), no other line is detected at a sufficiently high signal-to-noise to allow detailed constraints on its profile.

Because the line profiles observed cannot be well-modelled as Gaussians, we constrain the line profiles using a different methodology from that used for the LRIS data. We aim to constrain the velocity offset of the line peak, the upper FWHM, and the lower FWHM. First, we fit a second degree polynomial in the range $6560-6570\,{\rm \AA}$ of the pixel with the maximum flux. The line centroid is then the wavelength at the peak of the polynomial. We measure the uncertainties on this centroid by repeating this process 1000 times, where each time we randomize the spectrum based on the observed errors. This method of finding the line centroid is approximate: the centroid is not well defined for the unusual profiles observed here. We then measure the upper and lower FWHM as the distance to the pixels on either side of the peak at which the flux levels drop below half of the peak flux. The results are reported in Table~\ref{tab:lines}.

{\addtolength{\tabcolsep}{-2pt}   \begin{deluxetable*}{c | c | cccccc}
\centerwidetable
\tablewidth{10pt}
\tablecaption{Emission line fit parameters  \label{tab:lines}}
\tablehead{ \colhead{\multirow{2}{*}{Name}} & \colhead{\multirow{2}{*}{MJD}} & \colhead{\multirow{2}{*}{Line Name}} &  \colhead{$\Delta v$} & \colhead{$\sigma_{\rm FWHM, upper}$} & \colhead{$\sigma_{\rm FWHM, lower}$} & \colhead{$f_{\rm line}$} & \colhead{$L_{\rm line}$}\\
 \colhead{} &  \colhead{} &  \colhead{} & \colhead{km s$^{-1}$} & \colhead{km s$^{-1}$} & \colhead{km s$^{-1}$} & \colhead{$10^{-16}$ erg cm$^{-2}$ s$^{-1}$} & \colhead{$10^{39}$ erg s$^{-1}$}}
\startdata
\multirow{15}{*}{VT J1008} & \multirow{7}{*}{59676}  & [OII]$\lambda \lambda3726,3728$ & $99.9 \pm 42.3$ & $86.4 \pm 32.9$ & $-$ & $0.18 \pm 0.07$ & $0.09 \pm 0.03$\cr 
 &  & HeII$\lambda4685$ & $-18.6 \pm 21.7$ & $379.3 \pm 18.3$ & $-$ & $3.6 \pm 0.2$ & $1.7 \pm 0.1$\cr 
 &  & H$\beta$ & $205 \pm 36$ & $583 \pm 119$ & $1041 \pm 184$ & $3.1 \pm 0.9$ & $1.5 \pm 0.4$\cr 
 &  & [OIII]$\lambda5006$ & $30.2 \pm 19.5$ & $126.6 \pm 17.7$ & $-$ & $0.84 \pm 0.09$ & $0.40 \pm 0.05$\cr 
 &  & HeI$\lambda5875$ & $-72.5 \pm 45.4$ & $352.4 \pm 45.3$ & $-$ & $1.8 \pm 0.2$ & $0.8 \pm 0.1$\cr 
 &  & [FeX]$\lambda6374$ & $16.2 \pm 25.1$ & $342.9 \pm 27.0$ & $-$ & $2.6 \pm 0.2$ & $1.2 \pm 0.1$\cr 
 &  & H$\alpha$ & $140 \pm  6$ & $732 \pm 18$ & $1135 \pm 22$ & $27.7 \pm 0.5$ & $13.2 \pm 0.2$\cr\cline{2-8}
 & \multirow{7}{*}{59616}  & [OII]$\lambda \lambda3726,3728$ & $242.9 \pm 34.6$ & $99.1 \pm 25.1$ & $-$ & $0.31 \pm 0.08$ & $0.15 \pm 0.04$\cr 
 &  & HeII$\lambda4685$ & $52.0 \pm 34.5$ & $406.0 \pm 32.1$ & $-$ & $3.0 \pm 0.2$ & $1.4 \pm 0.1$\cr 
 &  & H$\beta$ & $115 \pm  3$ & $779 \pm  4$ & $779 \pm  4$ & $2.5 \pm 0.0$ & $1.19 \pm 0.01$\cr 
 &  & [OIII]$\lambda5006$ & $128.3 \pm 17.6$ & $114.8 \pm 18.2$ & $-$ & $0.8 \pm 0.1$ & $0.4 \pm 0.1$\cr 
 &  & HeI$\lambda5875$ & $130.5 \pm 59.0$ & $277.3 \pm 45.6$ & $-$ & $1.1 \pm 0.2$ & $0.5 \pm 0.1$\cr 
 &  & [FeX]$\lambda6374$ & $-29.6 \pm 29.4$ & $239.8 \pm 28.3$ & $-$ & $1.6 \pm 0.2$ & $0.8 \pm 0.1$\cr 
 &  & H$\alpha$ & $146 \pm  8$ & $732 \pm 26$ & $1030 \pm 34$ & $22.5 \pm 0.7$ & $10.7 \pm 0.4$\cr\cline{2-8}
& \multirow{1}{*}{59909} & H$\alpha$ & $1466.23 \pm 0.00$ & $7.09 \pm 0.00$ & $4.47 \pm 0.00$ & $-$ & $-$\cr\cline{2-8}
& \multirow{1}{*}{60029} & H$\alpha$ & $25.69 \pm 0.00$ & $372.28 \pm 0.00$ & $535.05 \pm 0.00$ & $-$ & $-$\cr\hline
\multirow{14}{*}{VT J2012} & \multirow{7}{*}{59464}  & [OII]$\lambda \lambda3726,3728$ & $-786.9 \pm 516.9$ & $821.8 \pm 486.9$ & $-$ & $0.5 \pm 0.3$ & $0.4 \pm 0.2$\cr 
 &  & HeII$\lambda4685$ & $531.3 \pm 92.6$ & $464.8 \pm 106.6$ & $-$ & $1.2 \pm 0.2$ & $0.8 \pm 0.1$\cr 
 &  & H$\beta$ & $694 \pm 39$ & $605 \pm 120$ & $607 \pm 120$ & $0.8 \pm 0.2$ & $0.6 \pm 0.1$\cr 
 &  & [OIII]$\lambda5006$ & $43.9 \pm 29.3$ & $106.3 \pm 20.9$ & $-$ & $0.41 \pm 0.08$ & $0.28 \pm 0.05$\cr 
 &  & HeI$\lambda5875$ & $694.9 \pm 69.1$ & $121.7 \pm 63.3$ & $-$ & $0.5 \pm 0.2$ & $0.4 \pm 0.1$\cr 
 &  & [FeX]$\lambda6374$ & $-120.0 \pm 370.1$ & $472.1 \pm 219.7$ & $-$ & $0.3 \pm 0.2$ & $0.2 \pm 0.1$\cr 
 &  & H$\alpha$ & $718 \pm 12$ & $729 \pm 33$ & $1083 \pm 79$ & $11.6 \pm 0.4$ & $7.8 \pm 0.3$\cr\cline{2-8}
 & \multirow{7}{*}{59677}  & [OII]$\lambda \lambda3726,3728$ & $-871.1 \pm 62.5$ & $133.1 \pm 51.0$ & $-$ & $0.4 \pm 0.2$ & $0.3 \pm 0.1$\cr 
 &  & HeII$\lambda4685$ & $196.1 \pm 72.2$ & $241.3 \pm 40.3$ & $-$ & $0.9 \pm 0.2$ & $0.6 \pm 0.1$\cr 
 &  & H$\beta$ & $804 \pm 169$ & $370 \pm 1067$ & $370 \pm 252$ & $0.6 \pm 0.1$ & $0.4 \pm 0.1$\cr 
 &  & [OIII]$\lambda5006$ & $35.4 \pm 63.7$ & $92.2 \pm 50.8$ & $-$ & $0.4 \pm 0.1$ & $0.3 \pm 0.1$\cr 
 &  & HeI$\lambda5875$ & $707.2 \pm 84.8$ & $557.7 \pm 83.2$ & $-$ & $2.2 \pm 0.3$ & $1.5 \pm 0.2$\cr 
 &  & [FeX]$\lambda6374$ & $-337.0 \pm 105.2$ & $907.3 \pm 72.1$ & $-$ & $3.2 \pm 0.3$ & $2.1 \pm 0.2$\cr 
 &  & H$\alpha$ & $687 \pm  9$ & $958 \pm 28$ & $1075 \pm 32$ & $15.0 \pm 0.7$ & $10.1 \pm 0.5$\cr\cline{2-8}
 & \multirow{1}{*}{59909} & H$\alpha$ & $814.88 \pm 7.82$ & $187.75 \pm 25.62$ & $218.94 \pm 56.37$ & $-$ & $-$\cr\hline
 \enddata
\centering
\end{deluxetable*}}

\subsection{Results}

In this section, we briefly summarize the results from our emission line analysis. We also perform some basic analysis to constrain the parameters of the emitting region, which uses simple approximations for the emission line parameters. While we do not expect the results of this analysis to be exact, we use them to gain a rough understanding of the physical conditions in the emitting region.

We first consider the recombination lines: the Balmer and Helium emission. Both H$\alpha$ and H$\beta$ are detected from VT J1008 and VT J2012 in all observations. The H$\alpha$ lines from both objects have luminosities of ${\sim}10^{40}$ erg s$^{-1}$ and they are brightening with time. The H$\beta$ lines are fainter, at ${\sim}10^{39}$ erg s$^{-1}$ and are (insignificantly) fading for VT J2012 and brightening for VT J1008. We will discuss the line profiles in more detail later in this section, but note here that the Balmer lines from both objects have widths ${\sim}1000$ km s$^{-1}$. 

He\,II$\lambda 4685$, and He\,I$\lambda 5875$ lines are strongly detected from VT J1008, and all have luminosities ${\sim}10^{39}$ erg s$^{-1}$ and FWHM${\sim}700$ km s$^{-1}$. The lines are brightening with time. The He\,II line is faintly detected from VT J2012 at a similar width and with a redshift consistent with that of the Balmer emission, although such a component is not detected in the second observation epoch. He\,I lines are not detected from VT J2012. 

Let us assume that these luminosities are entirely produced by recombination in a spherical region. We will discuss a wider range of models in Section~\ref{sec:discussion}. With this assumption, we can approximately estimate the mass and volume of the recombining material, following the methodology described in Chapter 13 of \cite{Osterbrock2006agna.book.....O}. The ionized mass is given by
\begin{equation}
    M_{\rm ion} = \frac{1.4 L_{\rm H\alpha} m_p}{1.15 n_p \alpha_{\rm H\alpha} h \nu_{\rm H\alpha}},
\end{equation}
where we have assumed a pure Hydrogen and Helium gas where the Helium density is a tenth of the Hydrogen density and the Helium is equally divided between its two ionization states. Assuming $L_{\rm H\alpha} = 10^{40}$ erg s$^{-1}$ and $\alpha_{\rm H\alpha} \sim 10^{-14}$ cm$^{3}$ s$^{-1}$, appropriate for case B recombination, we find an ionized gas mass $M_{\rm ion} \sim 3000\,M_\odot (10^4\,{\rm cm}^{-3}/n_e) \sim 0.3 \,M_\odot (10^9\,{\rm cm}^{-3}/n_e)$. Note that this analysis is not correct at very high densities ${\sim}10^9\,{\rm cm}^{-3}$, but we include the results to give a rough sense of the expected ionized gas mass. These mass constraints corresponds to radii $R \sim 10^{18}\,{\rm cm}(10^4\,{\rm cm}^{-3}/n_e)^{2/3} \sim 10^{15}\,{\rm cm}(10^4\,{\rm cm}^{-3}/n_e)^{2/3}$, assuming a filling factor ${\sim}1$. Based on the observed variation in the line luminosities and profiles on timescales $\lesssim 60$ days, we expect that the emitting region has a size $\lesssim 10^{17}$ cm, implying $n_e \gtrsim 10^5$ cm$^{-3}$. As we will discuss in detail later, if this gas is stellar debris from a TDE with a ${\sim}1\,M_\sun$ star, we require a high density $n_e\gtrsim 10^9\,{\rm cm}^{-3}$ in order that the ionized gas mass be smaller than the stellar mass. Even with that requirement, a large fraction of the stellar debris must be contained in this dense emitting region. This corresponds to an emitting radius $\lesssim 10^{15}$ cm. If the gas is not stellar debris, it could be from the galaxy circumnuclear medium, in which case lower densities are possible. 

We can further constrain the physical parameters of the Balmer emitting region using the Balmer decrement. These observed luminosities imply that the Balmer decrements from both transients are remarkably high and are increasing: that of VT J1008 is ${\sim}9$ during the both epochs and that of VT J2012 is $14.5$ in the first epoch and $25$ in the second epoch. Typical Balmer decrements from unextincted, low density, photoionized gas are ${\sim}3$. If we assume the Balmer emission is produced by recombination in a low density gas (e.g., $n_e\approx 10^{2-4}$ cm$^{-3}$), the high Balmer decrement implies strong extinction. Assuming photoionized gas with $T=10^4$ K and a low density $n_e\approx 10^2$ cm$^{-3}$, the color excess is related to the Balmer decrement as
\begin{equation}
    E(B-V) = 1.97 \log \bigg[ \frac{\rm H\alpha/H\beta}{2.86} \bigg]
\end{equation}
For VT J1008 (VT J2012), this implies $E(B-V)\approx 1 (1.5)$ mag. 

In addition to dust, high densities and radiative transfer effects can increase the Balmer decrement, although considering the latter is beyond the scope of this work. In AGN broad line regions, which have densities ${\gtrsim} 10^9$ cm$^{-3}$ the decrement is often observed to reach values ${\sim}6$. We believe it likely that high densities may contribute high Balmer decrements observed here based on the properties of the other observed emission lines. 

First, high densities are favored by the detection of the [Fe\,X] coronal line from VT J1008. Coronal lines are emission lines with extremely high ionization potentials $>100$ eV, and they are most often observed from AGN. However, coronal lines from AGN always have [Fe\,X]$\lambda 6375$ to [O\,III]$\lambda 5007$ ratios $\ll 1$. We observe a ratio $>2$, which is unprecedented for AGN. This high ratio placess VT J1008 in the class of extreme coronal line emitters, which we will discuss in the next section. Here, we use the detection of [Fe\,X] to constrain the physical parameters of the emitting region. Because the width of the [Fe\,X] line is similar to the Balmer widths, within a factor of a few, we expect that they are coming from ${\sim}$the same distance from the central SMBH, so it is plausible that the conditions of the coronal line emitting gas are similar to those of the Balmer emitting gas. We can constrain the density of the coronal line emitting gas using the lack of an [Fe\,VII] detection. Generally, the [Fe\,X] coronal line is accompanied by the detection of [Fe\,VII] transitions, which have a lower ionization potential, and so would be expected to be stronger. One way to suppress [Fe\,VII] is to invoke high density gas: the [Fe\,VII] critical density is ${\sim}10^7$ cm$^{-3}$, whereas the [Fe\,X] critical density is ${\sim}10^{10}$ cm$^{-3}$. Alternatively, the [Fe\,VII] could be suppressed if the ionizing SED is peaked above 250 eV, although we disfavor this possibility given the detection of low ionization potential lines like H$\alpha$ and He\,II with similar FWHM. Thus, the detection of [Fe\,X] without [Fe\,VII] suggests that there is high density gas in the intermediate-line emitting region. Since we now are confident that there is $\gtrsim 10^7$ cm$^{-3}$ gas in the vicinity of the Balmer line emitting region, given the similar line widths of the Balmer and coronal lines, it is plausible that the Balmer emitting region is similarly dense. Note that coronal lines are not detected from VT J2012; however, throughout this paper we will adopt the simplifying assumption that both events are produced by similar mechanisms, so we assume that the previous argument holds for VT J2012 and VT J1008.

A high density would also explain the lack of intermediate width Oxygen lines. We detect neither intermediate width [O\,II] nor [O\,III] lines from either VT J1008 and VT J2012. Given that the ionizing source is such that we observe both Balmer and [Fe\,X] lines, we would expect Oxygen lines to be detectable, given the intermediate ionization potentials of these transitions. The [O\,III] line has a critical density ${\sim}10^{7}$ cm$^{-3}$; thus, high densities would explain the lack of a detection.

Based on the previous arguments, we consider it very likely that the intermediate line emitting region has a density $\gtrsim 10^7$ cm$^{-3}$ and a radius $\lesssim 10^{16}$ cm.

We can further constrain the conditions of the line emitting region using the line profiles. We primarily consider the Balmer line profiles, because these lines are among the brightest. The Balmer line profiles from both objects are asymmetric. The Balmer emission from VT J1008 shows a slightly redshifted peak a long blue tail. VT J2012 shows similarly asymmetric H$\alpha$ emission. The observed H$\beta$ emission is narrower and more symmetric, although H$\beta$ is quite faint from VT J2012, so it is possible that the signal-to-noise is affecting our result. In contrast to the Balmer lines from VT J1008, those from VT J2012 are redshifted by ${\sim}700$ km s$^{-1}$, with no significant evolution in line centroid between epochs. 

From the later-time, high resolution spectra, we see that the line profile for VT J1008 shows a broad, flat top, with a long blue tail. In the ${\sim}120$ days between ESI observations, the line profile became slightly more peaked towards the red side. Note that the apparent brightening may not be real $-$ the flux is continuum-normalized, but the continuum level is expected to be different in the two observations because the slit orientations were different. In contrast to that of VT J1008, the line profile for VT J2012 is peaked and symmetric. 

Based on these line profiles, the emitting region must be aspherical and, in the case of VT J2012, flowing away from the observer. 

In summary, VT J1008 and VT J2012 show transient, intermediate width (${\sim}700-1000$ km s$^{-1}$) Balmer, He\,II, He\,I, and [Fe\,X] emission. Based on the observed luminosities and line profiles, we consider it likely that the emission is arising from a dense (${\gtrsim}10^7$ cm$^{-3}$), compact (${\lesssim}10^{16}$ cm), aspherical, outflowing emitting region. 

\subsection{Comparison to published TDE observations}

\begin{figure*}
    \centering
    \includegraphics[width=\textwidth]{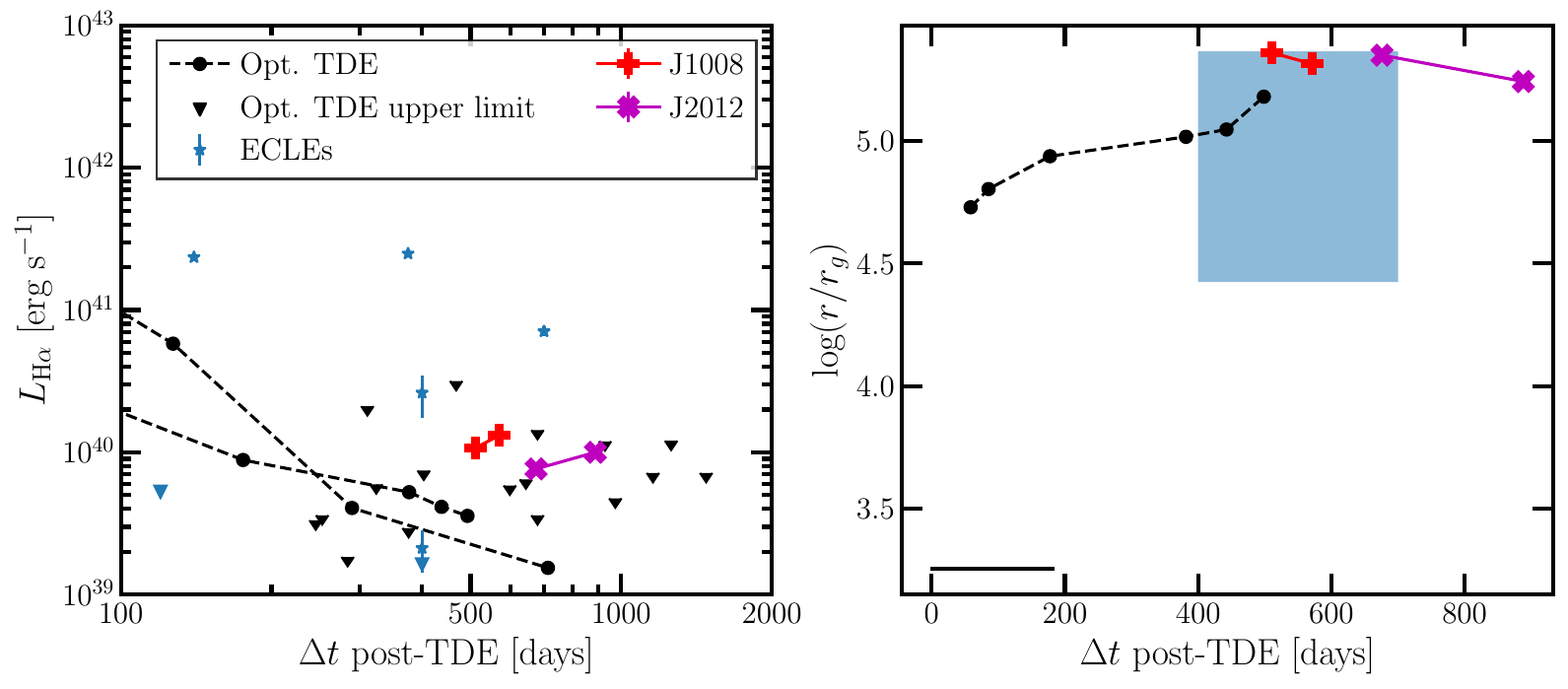}
    \caption{Comparison of the H$\alpha$ emission line from VT J1008 and VT J2012 to optically-selected TDEs (\citealp{Brown2017MNRAS.466.4904B, Hammerstein2022}, E. Hammerstein, private communication) and ECLEs \citep[][]{Wang2012ApJ...749..115W}. In the {\it left} panel, we show H$\alpha$ luminosity lightcurves for VT J1008 (red crosses), VT J2012 (magenta X's), ECLEs (blue stars), and optically-selected TDEs (black circles). The H$\alpha$ luminosities of VT J1008 and VT J2012 are much brighter at late times than those of optically-selected TDEs. They are more comparable to the ECLEs. In the \textit{right} panel, we show the distance of the emitting region from the central SMBH, $r$, implied from the H$\alpha$ width. The radio-selected TDEs are at larger radii than almost every TDE and ECLE. }
    \label{fig:Halpha_comp}
\end{figure*}

In the rest of this section, we compare the observed emission lines to observations of published transients. We first compare these objects to TDEs. In the next section, we discuss coronal-line emitting transients, some of which are TDEs and some of which are of ambiguous origin. Then, we compare to non-TDE transients. In this section, because we aim to compare to previous TDEs, we will focus on those lines that have been studied in published TDEs; namely, we focus on the Hydrogen and Helium lines.

We begin by discussing the Hydrogen and Helium lines. Optically-detected TDEs are well-established to produce these transient spectral features. These TDEs can be divided into four classes based on their early time ($\lesssim 6$ month post-optical flare): (1) Hydrogen rich TDEs, which should broad ${\sim}10^4$ km s$^{-1}$ Balmer features; (2) Hydrogen and Helium TDEs, which show ${\sim}10^4$ km s$^{-1}$ Balmer features and a complex of emission lines near He\,II$\lambda 4686$, typically including N\,III$\lambda 4640$ and N\,III $\lambda4100$; (3) Helium TDEs, which show only a broad (${\sim}10^4$ km s$^{-1}$) He\,II$\lambda 4686$ line; and, (4) featureless TDEs, which show no transient spectral features and have brighter optical flares and, typically, higher redshift host galaxies relative to those of the former three TDE classes. 

Our events would appear to most closely resemble the Hydrogen and Helium TDEs based on the detection of both Balmer and He\,II lines. However, there are many differences between our observed lines and those observed from Hydrogen and Helium TDEs. First, we do not observe N\,III lines. The N\,III lines are expected to be produced by the Bowen flourescence mechanism, which only operates under specific physical conditions. It is feasible that, at late times, it is not able to operate. 

Our observed lines also differ from typical TDEs because of their high luminosities and narrow widths. These are highlighted in Figure~\ref{fig:Halpha_comp}. The left panel of this figure shows the evolution of H$\alpha$ width and luminosity for a sample of optically-selected TDEs. The luminosities of our lines are both $\gtrsim 10^{40}$ erg s$^{-1}$, and are brightening with time ${\sim}2$ years post-TDE. Very few other optically-detected TDEs have detectable H$\alpha$ at such late times: most of the available $3\sigma$ upper limits are at or below the luminosity detected from our events. The TDEs ASASSN 14li and ASASSN 14ae both have H$\alpha$ detections ${\gtrsim}500$ days post-TDE, but the observed luminosities are at least a factor of a few fainter than those from our TDEs, and they both show declining H$\alpha$ luminosities whereas our events are brightening.

The right panel of Figure~\ref{fig:Halpha_comp} shows the gravitational radius in units of Schwarzschild radii $r_g$ of the H$\alpha$ emitting region for our TDEs, ASASSN-14li, and the typical early-time optically-selected TDE. We calculate this radius from the FWHM of the lines $v$ as $r/r_g = 2(v/c)^{2}$. For the typical early-time optically-selected TDE, we assume $v\sim 10^4$ km s$^{-1}$. The H$\alpha$ lines from our TDE are much broader than those from ASASSN 14li and those from early-time observations of optically-selected TDEs. Our TDEs have $\log r/r_g \sim 5.5$, and the radius is decreasing. The typical optically-selected TDE has $\log r/r_g \sim 3$. ASASSN 14li has $\log r/r_g \sim 5$ at ${\sim}500$ days post-TDE, although the radius is increasing and it is possible it evolved to match our events after the last observations. 

In summary, the lines from our TDEs most closely match those from the Hydrogen+Helium TDE class. However, they are significantly brighter and narrower than any previously observed TDE. 

\subsection{Comparison to coronal-line emitting transients}

There have been a few bona-fide TDEs and a growing sample of ambiguous transients with strong coronal line emission, like that observed from VT J1008. These objects are typically referred to as the extreme coronal line emitters (ECLEs). The first known ECLE, SDSS J0952+2143, was discovered by \cite{Komossa2008ApJ, Komossa2009} in a search for galaxies in SDSS with evolving spectral features. This source was associated with X-ray, optical, IR, and UV flares. There is no radio detection reported, although extensive radio follow-up was not performed. The source showed many transient emission lines, including bright coronal, He\,II and Balmer features. The coronal lines and He\,II lines are broad, with FWHM${\sim}800$ km s$^{-1}$. The Balmer lines were decomposed into multiple components: a narrow component consistent with the host galaxy dispersion and redshift, a redshifted broad component, and two unresolved horns on either side of the rest-frame component. The redshifted, broad H${\alpha}$component had a velocity offset of ${\sim}560$ km s$^{-1}$, a FWHM${\sim}1930$ km s$^{-1}$, and a luminosity ${\sim}10^{41}$ erg s$^{-1}$. The redshifted, broad H${\beta}$component had similar parameters, but with a luminosity implying a Balmer decrement ${\sim}9$. Both Balmer lines were first detected ${\sim}1$ year after the associated optical flare and remained bright ${\sim}3$-post optical flare, although they faded in that time. 

The cause of this transient emission is unknown. The event could be a TDE, but it also consistent with a supernova with extreme coronal line emission. AGN-like emission lines are detected from the host galaxy, so the emission could also be associated with a flaring AGN. If the source is caused by a TDE, \cite{Komossa2009} argues that the broad lines are likely produced by photoionized stellar debris that has become unbound and forms eccentric streams surrounding the SMBH. They argue that the unresolved, narrow horns are produced by shocks in a neutral medium, but \cite{Heng2010PASA...27...23H} shows that such a mechanism would require an unreasonable Hydrogen density.

The emission lines from SDSS J0952+2143 are remarkably similar to those observed from VT J1008 and VT J2012, although we do not detect narrow emission at the location of the Balmer lines and the Oxygen lines detected from our sources are much fainter than those from SDSS J0952+2143. We also do not detect multiwavelength flares analogous to those from SDSS J0952+2143. However, it is plausible that some of the differences between our sources and SDSS J0952+2143 could be reconciled by invoking a TDE in a galaxy with pre-existing accretion disk and/or a different line-of-sight.

After the discovery of SDSS J0952+2143, \cite{Wang2012ApJ...749..115W} performed a search for transient coronal-line emitting galaxies in SDSS. They identified a sample of seven non-active galaxies with strong coronal line emission. The host galaxies all had narrow line emission, with six qualifying as BPT H\,II galaxies and one bordering the LINER and H\,II regions. Four of the objects showed $\geq 3\sigma$ variations in their optical continua in the ${\sim}$months-years before the spectroscopic observations, measured by comparing their SDSS spectral and fiber magnitudes. In five of the seven sources, intermediate width emission lines are detected with FWHM ${\sim}880-2600$ km s$^{-1}$. The lines were fading in all objects. Broad He\,II$\lambda 4686$ is detected in three objects.

We overlay the luminosities and widths of the H$\alpha$ from these ECLEs on the panels in Figure~\ref{fig:Halpha_comp}. Note that the time since optical flare is very uncertain for these events. We adopted the time between the SDSS spectroscopic and photometric observations for those objects with detected optical flares, and 400 days for those objects without detected optical flares. The broad emission from these events much more closely resembles that from VT J1008 and VT J2012. The primary differences between these ECLEs and our events come from their host galaxies: the \cite{Wang2012ApJ...749..115W} ECLEs and SDSS J0852+2143 show strong nebular emission, whereas are events are in quiescent galaxies. Recent work, however, has suggested that, when considering the full ECLE population, they do tend to have TDE-like host galaxies; i.e., host galaxies that more closely resemble those of VT J1008 and VT J2012.

There is growing evidence that ECLEs are definitive TDEs, primarily based on arguments about the required ionizing flux, the optical light curve shapes for those few events with well-sampled light curves, and the similarities of TDE and ECLE host galaxies \citep[][]{Hinkle2023arXiv230305525H}. However, most of the known ECLEs are in galaxies with nebular emission lines from the host galaxies, rendering this conclusion uncertain: it is impossible to exclude an AGN origin for these events. We still lack a large sample of bona-fide TDEs with coronal line detections. It is, then, intriguing how closely the multiwavelength, transient properties of ECLEs resemble VT J1008 and VTJ2012, both of which have quiescent hosts, well-sampled optical light curves, and other multiwavelength data that allow us to argue that they are in fact bona-fide TDEs (see Section~\ref{sec:flare_origins}).

\subsection{Comparison to ambiguous and non-TDE transients}

In addition to TDEs and ECLEs, other transients can produce lines similar to those observed here. In particular, our objects resemble some supernovae. In particular, type IIn supernovae have been detected with similar line profiles. For example, SN2012ab is an optically-flaring Type IIn supernovae hosted near the nucleus a spiral galaxy that was discovered by \cite{Bilinski2018MNRAS.475.1104B}. SN2012ab is associated with an intermediate width H$\alpha$ component of width ${\sim}4500$ km s$^{-1}$ that is redshifted by ${\sim}800$ km s$^{-1}$. The intermediate width component was first detected ${\sim}7$ days after the optical flare, alongside a broad H$\alpha$ component with FWHM ${\sim}20000$ km s$^{-1}$. The intermediate width component is still detected ${\sim}1200$ days post-event. The late-time spectrum is not of sufficiently high signal-to-noise to constrain the presence of a late-time broad component.  \cite{Bilinski2018MNRAS.475.1104B} argues that SN2012ab is a type IIn supernova based on the observed spectral features and the optical light curve. The unusual H$\alpha$ emission is caused by interaction with an aspherical circumstellar material. They note that they cannot rule out a TDE origin; however, if the event is a TDE, it would be highly unusual because of the asymmetric material needed to produce the observed emission lines.

Based on the emission lines produced by VT J1008 and VT J2012, we see similar evidence for asphericities as observed for SN 2012ab. In contrast to SN 2012ab, VT J1008 and VT J2012 are hosted by quiescent galaxies. Moreover, no broad H$\alpha$ component is detected from our events, although our spectra were taken $>500$ days post-optical flare, and it is possible that a broad component was present at early times.

\subsection{Summary}

VT J1008 and VT J2012 are associated with Balmer, Helium, and coronal line emission. The line widths are ${\sim}1000$ km s$^{-1}$, which is much narrower than typical lines detected from optically-selected TDEs. Based on the line parameters, we suggested that the emission comes from a dense ${\gtrsim}10^7$ cm$^{-3}$ and compact ${\lesssim}10^{16}$ cm$^{-3}$ region. The mass of the ionized, recombination line-emitting gas is likely $\gtrsim 0.03-0.3\,M_\sun$; i.e., the mass is a large fraction of a solar mass. These emission lines do not resemble those detected from the optically-selected TDE sample. Instead, the observed emission lines most closely resemble those observed from the ambiguous class of ECLEs, which have been proposed, though not confirmed, to be TDEs.

\section{Discussion} \label{sec:discussion}

\subsection{Summary}

\begin{itemize}
    \item VT J1008 and VT J2012 are radio-selected transients in the nuclei quiescent, green valley host galaxies. The host galaxies have SMBHs with masses $M_{\rm BH} \sim 10^{5-6}\,M_\odot$. 
    \item VT J1008 and VT J2012 have optical counterparts, with lightcurves typical of optically-selected TDEs except that they are slightly fainter and cooler at peak. They do not have detectable X-ray or IR counterparts.
    \item VT J1008 and VT J2012 have radio counterparts with GHz luminosities ${\sim}10^{38}$ erg s$^{-1}$ that have SEDs consistent with wide-angle, low velocity $\beta \lesssim 0.1$ outflows at radii ${\sim}0.1$ pc.
    \item VT J1008 and VT J2012 both have transient, intermediate-width Balmer and He\,II emission detected ${\sim}2$ years post-TDE. Their H$\alpha$ luminosities are ${\sim}10^{40}$ erg s$^{-1}$ and are likely increasing. All the lines have FWHM ${\sim}700-1000$ km s$^{-1}$, and may be broadening slightly with time. VT J1008 also has strong He\,I and [Fe\,X] emission with similar line widths. 
    \item The observed transient lines detected from these radio-selected events are reminiscent of Hydrogen+Helium TDEs. However, the lines are much more luminous at late-times than other optically-detected TDEs, and they are narrower than is typically observed from optically-selected TDEs. 
    \item Instead, the observed spectral features more closely resemble the emission lines associated with some extreme coronal line emitters. A subset of these objects are known to have intermediate width lines, in some cases with centroids that are redshifted relative to the host (e.g., J0952+2143).
\end{itemize}

\subsection{Are VT J1008 and VT J2012 tidal disruption events?} \label{sec:flare_origins}

Before delving into the origin of the transient emission lines, we briefly consider and rule out the possibility that VT J1008 and VT J2012 are not tidal disruption events. The three most likely origins for these events are: stellar explosions, active galactic nuclei flares, or TDEs. In the following, we consider each of these possibilities. We assume both events are caused by the same type of event, which is justified given the strong similarities in their multiwavelength properties and host galaxies.

\noindent \textbf{Stellar explosions.}

We first consider the possibility that these events are stellar explosions. Supernovae can produce radio-loud events with optical flares and transient optical spectral lines. However, the only type of supernovae that has been observed to produce radio outflows with velocities ${\sim}0.1c$, as observed for our events, are SN Ic-BL, so we only consider this type of event to be possible. There are a number of factors that make VT J1008 and VT J2012 unusual for SNe Ic-BLs. First, these events are in the nuclei of their host galaxies: their offsets in units of the host half-light radii are consistent with zero. SNe Ic-BLs tend to lie in the outskirts of their host galaxies, with a median offset relative to host half-light radius of $0.7 \pm 0.2$ \citep[][]{Japelj2018}. Second, our events are hosted by non-star forming galaxies, whereas SNe Ic-BL hosts tend to be star-forming ($0.3\%$ of core-collapse supernovae hosts are quiescent). 

The observed optical light curve is very unusual for SN Ic-BL: these SN often, though not always, show rapid post-peak cooling, which is not present for our events \citep[][]{Taddia2019}. While the late time spectra of SN Ic-BL are not well constrained, intermediate width features such as we see are unprecedented (S. Anand, private communication). 

While we cannot definitively rule out a SN Ic-BL origin, there are many factors that would make our events extremely unusual. Hence, we do not believe our events are associated with stellar explosions.

\noindent \textbf{Active galactic nuclei flares.}

We next consider active galactic nuclei flares. AGN can produce variability and flaring across the electromagnetic timescale and over a wide range of timescales. The main evidence against an AGN origin for VT J1008 and VT J2012 is the lack of any evidence that there was active accretion prior to the optical flare. The optical spectra do not show any evidence for strong AGN emission. It is possible that the weak [O\,III] detections are caused with an AGN, but it would imply a very weak AGN. Moreover, that line could also be caused by weak star formation. The IR colors of the host galaxy also show no evidence for AGN activity; likewise, the lack of an X-ray detection and the lack of optical variability support the hypothesis that these are quiescent galaxies.

From the above evidence, we disfavor an AGN origin. Of course, none of these arguments completely rule out the possibility of a very weak, flaring AGN. In this case, some trigger has caused a large amount of mass to be dumped on the central SMBH. Such events are extremely poorly understood: it is not clear whether such events could even happen without a discrete object like a star venturing near the SMBH. However, given the broad consistency of our events with TDEs and the large complications associated with interpreting low luminosity AGN flares, we do not consider this possibility further.

\noindent \textbf{Tidal disruption event.} 

After ruling out AGN flares and stellar explosions, we are left with TDEs. As we have discussed in this work and in Paper I, the host galaxy, optical lightcurves, and radio emission of VT J1008 and VT J2012 are broadly consistent with optically-selected TDEs. Henceforth, we consider these events to be definitive TDEs. 

\subsection{The origin of the transient spectral features} \label{sec:origin}

In the rest of this paper, we discuss the origin of the transient spectral features. We make the strong but necessary assumptions that (1) the ionization source producing the Balmer emission is the same that produces the higher ionization lines and (2) the same emission mechanism is active in both VT J1008 and VT J2012. 

\subsubsection{Are the lines associated with a shock?}

\begin{figure*}
    \centering
    \includegraphics[width=\textwidth]{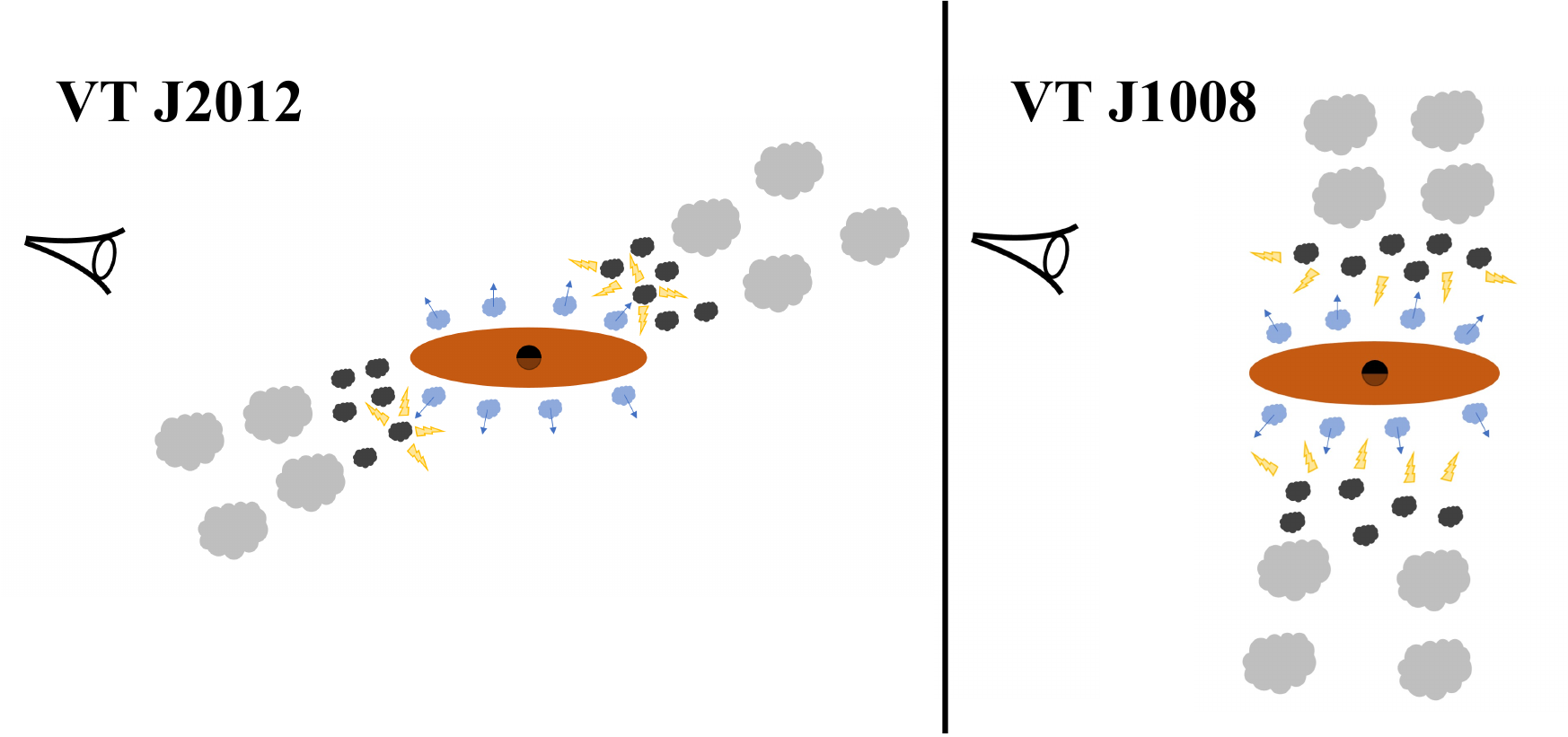}
    \caption{An example cartoon geometry that could produce either redshifted emission without a blueshifted counterpart, or non-shifted emission. We invoke a dusty torus that is misaligned from the TDE-produced accretion disk. The blue clouds represent accretion disk winds, the black clouds represent dense gas in the torus, and the grey clouds represent less dense gas at larger distances. When the disk winds slam into the dense torus, radiative shocks are produced. The resulting free-free emission photoionizes gas in the vicinity, including outflow gas launched from the disk, producing the observed emission lines. In the {\it left} panel, the blueshifted component is obscured from the observer by the dusty torus. In the {\it right} panel, no redshifted or blueshifted components are produced. }
    \label{fig:geometry}
\end{figure*}

In both VT J1008 and VT J2012, we know that fast outflows exist given the observed radio emission. It is not a far stretch to imagine that the transient emission lines are also associated with a shocking outflow. Shocks can produce emission lines in multiple ways, depending on the shock velocity and medium density. If the shocked gas can cool efficiently, it will produce strong free-free emission that can photoionize the surrounding medium. Otherwise, the shock is ``nonradiative'' and primarily emits through collisional ionization.

We can determine if a shock is radiative by comparing the age to the cooling time. To compute the cooling time, we need to know the density $n$ and the shock velocity $v_s$. From the shock velocity, we first must compute the shock temperature $T_s$: 
\begin{gather}
    T_s = \frac{2(\gamma-1)}{(\gamma+1)^2} \frac{m_p}{k_B} v_s^2  = 2.2 \times 10^7\,{\rm K} \, \bigg( \frac{v_s}{10^3\,{\rm km\,s}^{-1}} \bigg)^2,
\end{gather}
where $\gamma$ is the adiabatic index and we adopt $\gamma=5/3$. We consider shock velocities in the range ${\sim}500-10^4$ km s$^{-1}$, which approximately match the observed line widths. Slow shocks $v_s \lesssim 100$ km s$^{-1}$ do not produce ${\sim}1000$ km s$^{-1}$ emission lines.

Then, we must compute the cooling rate $\Lambda(T)$, which, following \cite{Draine2011_book}, we approximate as 
\begin{gather}
    \frac{\Lambda(T)}{{\rm erg\,cm}^{3}\,{\rm s}^{-1}} = \begin{cases}
    2.3 \times 10^{-24}\,  \big(\frac{T_s}{10^6\,{\rm K}}\big)^{0.5}, & T_s > 10^{7.3}\,{\rm K} \\
    1.1 \times 10^{-22}\, \big(\frac{T_s}{10^6\,{\rm K}}\big)^{-0.7}, & 10^{5}\,{\rm K} < T_s \le 10^{7.3}\,{\rm K}.
    \end{cases}\label{eq:cooling}
\end{gather}
At high densities, such as those we will consider, the cooling rate can be suppressed because collisional de-excitation reduces cooling through heavy elements. However, at the high temperatures we are consider $\gtrsim 10^7$ K, cooling through heavy elements in subdominant and this suppression is minimal \citep[][]{Wang2014MNRAS}, so we do not consider it further.

With these definitions, the cooling time is given by
\begin{gather}
    t_{\rm cool} = \frac{3 k_B T_s}{n \Lambda(T)} \nonumber \\
    = 1.3\,{\rm years} \, \bigg( \frac{T_s}{10^7\,{\rm K}} \bigg) \bigg( \frac{n}{10^6\,{\rm cm}^{-3}} \bigg)^{-1} \bigg( \frac{\Lambda(T)}{10^{-22}\,{\rm erg\,cm}^{3}\,{\rm s}^{-1}} \bigg)^{-1}.
\end{gather}
If the shock has $t_{\rm age} \approx 1\,{\rm year}$, then the shock is non-radiative if the density $n \lesssim 10^7$ cm$^{-3}$ and $v_s \sim 10^3$ km s$^{-1}$, or if the shock is near-relativistic and has $10^7 \lesssim n/{\rm cm}^{-3} \lesssim 10^8$.

With these constraints on the parameter space, we now explore possible emission mechanisms for both radiative and non-radiative shocks. First, we consider a non-radiative shock. In this case, we know that the density is relatively low ($n \lesssim 10^7$ cm$^{-3}$) if the shock is non-relativistic. One possible emission mechanism comes from an analogy to type IIn supernova, which produce similar velocity shocks but in such low density material ($n\sim0.1-10$ cm$^{-3}$). In these events, Balmer emission resembling that which we observe is produce from Balmer dominated shocks, which occur when the pre-shock material is partly neutral. In this case, we can approximate the mass of H$\alpha$ emitting atoms as:
\begin{gather}
    M_{\rm H\alpha} = \frac{L_{\rm H\alpha} t_{\rm H\alpha} m_{\rm H}}{E_{\rm H\alpha} \epsilon_{\rm H\alpha}}\nonumber \\
    = 438 M_\odot \frac{L_{\rm H\alpha}}{10^{40}\,{\rm erg\,s}^{-1}} \frac{t_{\rm H\alpha}}{1\,{\rm year}}.
\end{gather}
Here, $L_{\rm H\alpha}$ is the average H$\alpha$ luminosity and $t_{\rm H\alpha}$ is the duration of the H$\alpha$ emission. We adopted fiducial values $L_{\rm H\alpha}\sim 10^{39}$ erg s$^{-1}$ and $t_{\rm H\alpha}\sim 1$ year, which are likely correct to within an order of magnitude. $m_{\rm H}$ is the mass of a neutral Hydrogen atom, $E_{\rm H\alpha} \sim 2$ eV is the energy of an $H\alpha$ photon, and $\epsilon_{\rm H\alpha} \sim 0.2$ is the fraction of excited Hydrogen atom that undergo the H$\alpha$ transition. If the density of the emitting region is ${\lesssim} 10^8$ cm$^{-3}$, then the size of the emitting region must be $\gtrsim 0.03$ pc. Assuming a source age of ${\sim}3$ years, the outflow that produced the Balmer dominated shocks must have a velocity $v \gtrsim 10^5$ km s$^{-1}$. This value is inconsistent with the width of the observed lines, which is expected to correspond to roughly the velocity of the shock for Balmer dominated shocks. Because of this inconsistency, we do not consider a Balmer dominated shock a feasible cause of the observed emission. However, most modelling of Balmer dominated shocks, on which we base our discussion, assumes low densities applicable to supernova remnants. It is possible that similar shocks in high density environments could have different properties, in which case something similar to a Balmer dominated shock could produce the observed emission. Further exploration of this is beyond the scope of this work.

We now move to the possibility of a radiative shock. In this case, the shock has caused surrounding gas to heat up to ${\sim}10^7$ K as described earlier, which is cooling quickly via free-free emission. This emission ionizes the surrounding gas, which is likely at a similar density to the shocked, cooling gas ($n\gtrsim10^8$ cm$^{-3}$). From Equation~\ref{eq:cooling}, the cooling rate for gas at a temperature ${\gtrsim}2\times 10^7$ K is $\Gamma(T) \gtrsim 10^{-23}$ erg cm$^3$ s$^{-1}$. Assuming $n_e \gtrsim 10^7$ cm$^{-3}$ in a region of radius $r\sim0.01$ pc, we have a total luminosity of $\gtrsim 10^{41}$ erg s$^{-1}$. Most of these photons will be Hydrogen- and Helium-ionizing given the high temperature of the free-free emitting region.

In typical models of the emission lines produced by radiative shocks, Oxygen lines (e.g., \citealp{Allen2008ApJS..178...20A}) are produced. In this case, we would not expect to see Oxygen lines given the high density of the material and the low critical densities of the typical lines. Instead, we would only expect to see recombination lines, such as the observed Balmer and Helium emission, and lines with very high critical densities, such as [Fe\,X]. The detailed modelling required to constrain this possibility quantitatively is beyond the scope of this paper, but from this qualitative discussion we believe it feasible that the observed emission could be produced by a radiative shock. 

Now that we have established that the observed lines could be produced by a radiative shock, we must consider the centroid offsets from the host redshifts. Discussion of the line profiles is beyond the scope of this work. The intermediate width lines from VT J2012 are redshifted relative to the host galaxy, whereas the lines from VT J1008 are near the host redshift. There are two possible explanations for the line offsets: (1) the SMBH that produces VT J2012 is moving or (2) the emitting gas is outflowing. We are pursuing follow-up to constrain (1) and will consider it in a future paper; here, we consider (2). While it is possible that we do not detect a blueshifted event by chance, with a shock model it is feasible to produce a geometry where only redshifted velocities are possible.

In Figure~\ref{fig:geometry}, we show a cartoon of one model that can produce the observed lines (not to scale). First, we suppose that the TDE has produced an accretion disk that is producing disk winds \citep[see][for possible disk wind launch models]{winds}. These winds are outflowing, and collide with the circumnuclear medium (CNM) of the galaxy to produce the radiative shocks. While the structure of the CNM is poorly constrained, it is feasible that it is in a torus (or some extended, axisymmetric) structure, as is known to exist in AGN host galaxies. There is no a prior reason that this axisymmetric dust structure need be aligned with the TDE accretion disk: if both the disk  and torus orientations are set by the SMBH spin direction, gravitational precession wculd cause it to change relative to when the dusty torus was formed. If the disk orientation is related to the orbit of the disrupted star, it will be independent of the torus orientation. In Figure~\ref{fig:geometry}, we show two possible orientations. In the left panel, the torus is inclined relative to the disk, and in the right panel, the disk and torus orientation are perpendicular to each other. In the left panel, the disk wind clouds will tend to collide with the edge of the dusty structure. The clouds that are outflowing away from the observer (redshifted) will be visible, but those flowing towards the observer (blueshifted) are seen through a large column of dust. In the right panel, all of the clouds are visible, but no blueshift or redshift is expected. While this is a cartoon model, and it is unclear whether this dust structure is expected, it could reproduce the observed geometry.

In summary, a radiative shock model invoking an axisymmetric dusty structure misaligned with the TDE accretion disk could reproduce the observed lines, including their widths and offset from host redshift.

\subsubsection{Are the lines photoionized by a central source?}

If the lines are not associated with the shock, they are likely photoionized by a central source associated with the accretion induced by the TDE, as is observed at early times. The observed lines would thus be the evolved version of the early time lines observed from optically-selected TDEs. This leads to two questions: (1) can the models that produce the early time TDE emission also explain the late-time emission, and (2) why do we detect these lines in these radio-detected, optically-selected TDEs, when they do not seem to be present in most optically-selected TDEs?

The evolution of early-time TDE transient lines is not well explored, but available observations suggest that they tend to fade within ${\sim}1$ year. Before this work, none had been detected above $10^{40}$ erg s$^{-1}$ at times ${\gtrsim}1$ year post-optical peak. The only strong detection was from the extensively studied radio-emitting TDE ASASSN 14li, which had an H$\alpha$ detection ${\sim}1.5$ years post-optical peak at a luminosity $7\times 10^{38}$ erg s$^{-1}$.

The origin of these early-time lines is still debated. \cite{Roth2018ApJ...855...54R} presented a model where the lines originate from an extended, optically thick envelope surrounding the SMBH. The envelope reprocesses soft X-ray photons emitted during the accretion of the stellar debris. A percentage of the reprocessed emission produces an optical continuum, corresponding to the observed optical flares. Hydrogen and helium in the envelope become ionized and produce the observed emission lines, but because the envelope is optically thick, the Balmer lines are suppressed relative to the Helium emission. The resulting line profiles were analyzed by \cite{Roth2018ApJ...855...54R}, who how that an optically-thick envelope will produce ${\sim}10^4$ km s$^{-1}$ emission lines due to electron scattering, without requiring high velocity dispersion gas. If the optically-thick envelope is outflowing, the line profile will not be Gaussian but instead will have a blueshifted peak with an extended red wing. With time, the emission lines will narrow as the density decreases and electron scattering reduces. The time evolution of the line luminosities relative to the optical continuum level has not been explored in depth.

Narrower lines may be expected at late times, if the \cite{Roth2018ApJ...855...54R} model is correct. The line profiles that we observe, however, do not match those predicted by \cite{Roth2018ApJ...855...54R}. In the case of stationary gas a symmetric line profile is expected whereas outflowing gas would produce a blueshifted peak with a redshifted tail. In the observations of both VT J1008 and VT J2012, we see a redshifted peak with a blueshifted tail. It is possible that altering the geometry of the envelope could produce the observed emission, but the required modelling is beyond the scope of this work.

Another challenging aspect of a model where the lines are produced by photoionization from a central source is the association with radio emission. There is no reason, a-priori, that we would expect the intermediate width lines to preferentially occur in radio-emitting systems if they are produced by photoionization. One possible explanation is that both these lines and radio-emitting shocks are produced by TDEs with slow accretion rate decays, in which case the photoionizing continuum can remain sufficiently strong at late times to produce the observed emission lines. This explanation is subject to significant theoretical uncertainties; in particular, there is no expectation that events with slower accretion rate decays will tend to cause radio-emission. We also have no direct observational evidence that long-lived TDEs tend to produce radio emission. In Paper I, we saw no significant correlation between the decay of the optical light curve and the presence of radio emission. We do not detect any remarkable X-ray emission from these sources.

Alternatively, we can invoke a gas-rich environment to explain both the radio emission and the spectral features. Such an environment has been invoked for coronal line emitters in the past. Then, the radio detections are caused by shocks in the gas. While this is plausible, it is unclear why this would produce the unusual line profiles that are observed, nor why the emission would be entirely redshifted in the case of VT J2012. Unlike in the case of shock ionization, there is no clear geometry that could produce the observed redshift, unless the SMBH is recoiling or in a binary, which we will constrain in a future paper. This tension becomes stronger if we include events like J0952+2143, which also shows redshifted intermediate width emission. It seems improbable, though not impossible, both of these are TDEs by recoiling or binary SMBHs. Instead, we require dense, rapidly outflowing, asymmetrically-distributed photoionized gas. The gas could be the outflowing stellar debris, but, as we discussed in Section~\ref{sec:lines}, the mass of the ionized gas is a large fraction of a solar mass. Unless these events were caused by the discussion of high mass ($\gtrsim$a few solar mass) stars, we require that most of the unbound debris remains in a compact region. This is not expected based on current TDE theory.

In summary, while photoionization from a central source is a feasible model, we prefer a shock ionized model. It is possible that detailed models of the evolution of the transient emission lines from TDEs and the dust geometry and kinematics in the circumnuclear medium could reproduce the observed emission, but we do not currently have strong evidence to favor this model.

\section{Conclusions}

We have presented the multiwavelength properties of two radio-selected TDEs. The TDEs were selected from our sample of six radio-selected, optically-detected TDEs from the VLA Sky Survey. They were the only two TDEs in quiescent galaxies, and they showed unusual, intermediate-width Balmer and Helium emission. These events were otherwise fully consistent with optically-selected TDEs. We discussed the origin of the intermediate width emission lines in detail, and argued that they likely originate from a radiative shock. Alternatively, the lines could originate from outflowing, asymmetric, dense gas in the circumnuclear medium that is photoionized by the TDE, but we marginally disfavor this model.

One of the most intringuing findings in this work is that the transient spectral features observed from these two radio-selected TDEs share many characteristics with those from the ambiguous class of coronal line emitting transients, the ECLEs. This connection provides yet more evidence that ECLEs are caused by TDEs. Moreover, just as early time transient spectral features from TDEs allow the events to be subdivided into classes with different properties, these late-time spectral features may allow for a new TDE classification system: featureless late-time spectra, the extreme coronal line emitters with intermediate width recombination lines, and the extreme coronal line emitters without intermediate width recombination lines. These different classes likely correspond to physically different events; e.g., events with different amounts of circumnuclear material and/or those that do and do not allow for the launch of outflows. In future work, we hope to obtain late-time spectra for a large sample of TDEs, with the aim of further developing classification system and pinning down the physical causes of the late-time emission.

\bibliography{sample631}{}
\bibliographystyle{aasjournal}

\end{document}